# Electric-field control of the perpendicular magnetization switching in ferroelectric/ferrimagnet heterostructures

Pengfei Liu[1†], Tao Xu[2†], Qi Liu[3], Juncai Dong[4], Ting Lin[5], Qinhua Zhang[5], Xiukai Lan[1], Yu Sheng[1], Chunyu Wang[6], Jiajing Pei[4], Hongxin Yang[2], Lin Gu[7]*, Kaiyou Wang[1,8] *

[1]*State Key Laboratory for Superlattices and Microstructures, Institute of Semiconductors, Chinese Academy of Sciences, Beijing 100083, China*

[2]*Ningbo Institute of Materials Technology and Engineering, Chinese Academy of Sciences, Ningbo 315201, China*

[3]*Department of Physics, Southern University of Science and Technology, Shenzhen, 518055, China*

[4]*Beijing Synchrotron Radiation Facility, Institute of High Energy Physics, Chinese Academy of Sciences, Beijing, 100049, China*

[5]*Beijing National Laboratory for Condensed Matter Physics, Institute of Physics, Chinese Academy of Sciences, Beijing 100190, China*

[6]*Key Laboratory of Pressure Systems and Safety Ministry of Education, East China University of Science and Technology, Shanghai 200237, China*

[7]*Beijing National Center for Electron Microscopy and Laboratory of Advanced Materials, Department of Materials Science and Engineering, Tsinghua University, Beijing 100084, China*

[8]*Center of Materials Science and Optoelectronics Engineering, University of Chinese Academy of Sciences, Beijing 100049, China*

Electric field control of the magnetic state in ferrimagnets holds great promise for developing spintronic devices due to low power consumption. Here, we demonstrate a non-volatile reversal of perpendicular net magnetization in a ferrimagnet by manipulating the electric-field driven polarization within the Pb $(Zr_{0.2}Ti_{0.8})$ $O_3$ (PZT)/CoGd heterostructure. Electron energy loss spectra and X-ray absorption spectrum directly verify that the oxygen ion migration at the PZT/CoGd interface associated with reversing the polarization causes the enhanced/reduced oxidation in CoGd. Ab initio calculations further substantiate that the migrated oxygen ions can modulate the relative magnetization of Co/Gd sublattices, facilitating perpendicular net magnetization switching. Our findings offer an approach to effectively control ferrimagnetic net magnetization, holding significant implications for ferrimagnetic spintronic applications.

## 1. Introduction

The manipulation of magnetic states through electrical means, including electric-field and electric-current methods, paves the way for the development of efficient, fast, and non-volatile information storage and processing technologies [1–14]. Ferrimagnets, combining the strengths of ferromagnets and antiferromagnets, offer distinct advantages including facile detection of net magnetization by an external field, minimal stray field [15] and sub-terahertz spin dynamics [16–18]. These characteristics hold promise for the creation of high-density, picosecond-speed, and low-power devices [19,20]. Compensated ferrimagnet, containing a series of rare earth (RE)-transition metal (TM) compounds, where the spins of two inequivalent magnetic sublattices are coupled antiferromagnetically [21,22]. Electrically control the net magnetization switching in compensated ferrimagnets typically involves two approaches: one involves the simultaneous reversal of all sublattice magnetic moments [23]; while the other, alters the relative magnitude of magnetization among sublattices, such as using hydrogen injection [24], piezoelectric control [25] and high-k dielectric $HfO_2$ gating [26], as the net magnetization is primarily dominated by those with larger magnitudes [27].

Due to the advantages of non-volatility, reversibility and low-voltage, the influence of ferroelectric polarization on ferromagnetism has been extensively investigated [28–30]. Besides electric-field induced stress associated with ferroelectric polarization switching [31–33], the ferromagnetic states can also be controlled by the ferroelectric polarization induced orbital reconstruction [34], charge accumulation [35,36], spin transmission modulation [37] and current density gradient [38] etc. However, the ferroelectric/ferrimagnet heterostructure has not been investigated yet. Therefore, the ferroelectric polarization effect on ferrimagnetic states and the underlying physics attract considerable scientific and technological interest.

In this work, we demonstrate the reversible net magnetization switching of ferrimagnet CoGd controlled by the polarization of Pb $(Zr_{0.2}Ti_{0.8})$ $O_3$ (PZT) in PZT/CoGd/Pt hybrid heterostructure near the magnetic compensation temperature ($T_M$). With reversing the polarity of the ferroelectric PZT, the oxygen ion migrations induced by electric-field driven polarization can be observed at the ferroelectric/ferrimagnetic interface at an atomic-level scale. Combing with Electron energy loss spectra (EELS), X-ray absorption spectrum (XAS), and the first-principle calculations, we find the these migrations at the CoGd/PZT interface is responsible for this reversible magnetization switching.

## 2. Results and Discussion

### 2.1 Characterizations of the PZT/CoGd/Pt heterostructure

Epitaxial $Ca_{0.96}Ce_{0.04}MnO_3$ (CCMO, 20 nm, as a bottom electrode)/PZT (60 nm) bilayers were prepared on $SrTiO_3$(STO) substrate using a pulsed laser deposition technique. Then $Co_{77}Gd_{23}$ (4 nm)/Pt (5 nm) were deposited by magnetron sputtering. The growth and fabrication details are shown in **Experimental Section**. The Atomic Force Microscopy image, X-ray diffraction pattern, and Reciprocal Space Mapping of CCMO/PZT bilayers are shown in Supporting Information S1. These results confirm the high quality of our ferroelectric bilayer.

We then investigate the magnetoelectric transport of the Hall bar device of PZT/$Co_{77}Gd_{23}$ (4 nm)/Pt (5 nm), which is shown in Figure 1(a). The hysteresis of anomalous Hall effect (AHE) for the heterostructure at different temperatures are shown in Figure 1(b) and Figure S2. In the temperature range of 180~230 K, the AHE loops show anticlockwise from 180 to 208 K, whereas they become clockwise at 210 K and above. Meanwhile, with increasing the temperature, the magnitude of the AHE change $|\Delta R_{AHE}|$ decreases from 0.435 Ω at 180 K to 0.038 Ω at 208 K and recovers to a significant resistance of 0.430 Ω at 230 K. The AHE sign reversal crossing the $T_M$ ( ~ 209 K) is because the chirality of AHE (clockwise or anti-clockwise) is sensitive to the Co magnetization direction, and the Co magnetization is antiparallel to the orientation of $H_Z$ when the magnetization is dominated by Gd sublattice. Therefore, the net magnetization is Co-dominant above 210 K and Gd-dominant below 208 K, respectively. The magnitude of $\Delta R_{AHE}$ is still proportional to the perpendicular net magnetization [39], which is consistent with conventional compensated ferrimagnets. The X-ray magnetic circular dichroism (XMCD) spectra of Co $L_{2,3}$ and Gd $M_{4,5}$ at 100 K (below $T_M$) and 240 K (above $T_M$) were measured under fixed out-of-plane magnetic field of 1T to confirm the dominant sublattices. As shown in Figure 1(c) and (d), the signs of Co and Gd are opposite at the same temperature, demonstrating the antiferromagnetic coupling between Co and Gd sublattice. And the signs of both Co and Gd XMCD signals invert from 100 K to 240 K, indicating the dominant sublattice switched from Gd to Co.

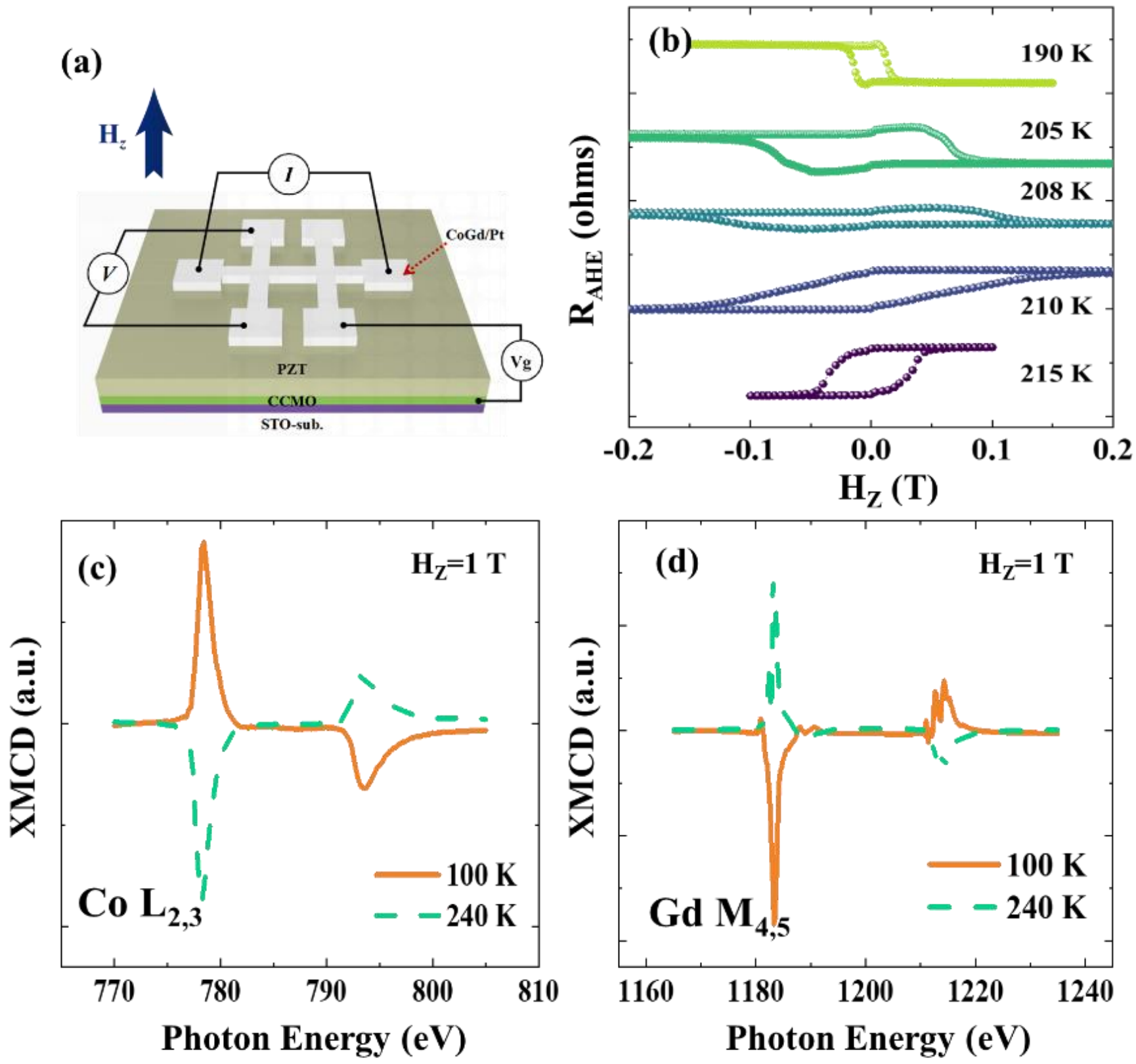

Figure 1 (a) The schematic diagram of the Hall bar structure of the CCMO/PZT/CoGd/Pt heterostructure. (b) The AHE resistance of the heterostructure as a function of the perpendicular field for several temperatures. Exemplary (c) and (d) Co $L_{2,3}$ and Gd $M_{4,5}$ XMCD spectra at 100 K (Gd dominance) and 240 K (Co dominance), respectively. The XMCD measurements were carried out under the fixed out-of-plane magnetic field of 1 T.

### 2.2 Electric-field control AHE by ferroelectric polarization

We further investigate how the polarization affects the magnetization switching of the ferrimagnet layer. The out-of-plane piezoelectric force microscopy was applied to check the polarization switching of the CCMO/PZT bilayer. With writing voltage -/+4.2 V, the dark (bright) square indicates the downward (upward) polarization in the center (peripheral) square (Figure 2(a)), indicating the full switching of polarization of PZT. The AHE measurements of the Hall device were performed at different temperatures for opposite polarizations (Figure 2 and S3(a) and (b)). The up/down polarization of the ferroelectric was set up by applying the pulsed gate voltage of +/-4.2 V with width of 100 μs from the CCMO bottom electrode to the CoGd/Pt layers (Figure 1(a)). At the temperature of 200 K below $T_M$ (Figure 2(b)), the hysteresis loop of AHE keeps anticlockwise by reversing the polarization direction, suggesting the Gd dominates at both polarizations. Although the dominant sublattices cannot be converted, the $H_C$ is increased by 231% from upward to downward polarization. At the temperature of 220 K, which is about 10 K above $T_M$ (Figure 2(c)), the hysteresis loops for both up- and downward polarization keep clockwise, indicating the Co dominance for both polarizations. However, the $H_C$ is decreased by 54% with polarization reversed from upward to downward. Although the polarization can effectively modulate the PMA at these two temperatures, but the changing trend is opposite. The change of $\Delta H_C$ in opposite polarizations at different temperatures away the $T_M$ point was demonstrated in the Supporting Information S3(c).

Interestingly, close to the $T_M$ (~ 210 K), the hysteresis loop is converted from clockwise to anticlockwise with polarization changed from downward to upward. The anticlockwise loop can be reversed back to clockwise by switching the polarization back downward (Figure 2(d)). The reversible hysteresis loops confirm that the dominant sublattice switching of ferrimagnet CoGd close to $T_M$ can be controlled by changing the polarization, demonstrating the switching of the net magnetization. To ensure the effective control of the magnetization by electric field, smaller Hall crosses of 30 μm width were fabricated (Supporting Information S4) to measure the $R_{AHE}$ with periodically changed the pulsed electric fields. With applied the external magnetic field of -/+ 0.15 T (Figure 2e), the $R_{AHE}$ periodically changes from negative/positive to positive/negative with changing the polarization from downward to upward, respectively. Therefore, the non-volatile perpendicular net magnetization switching of ferrimagnet CoGd can be achieved by electrical-field control.

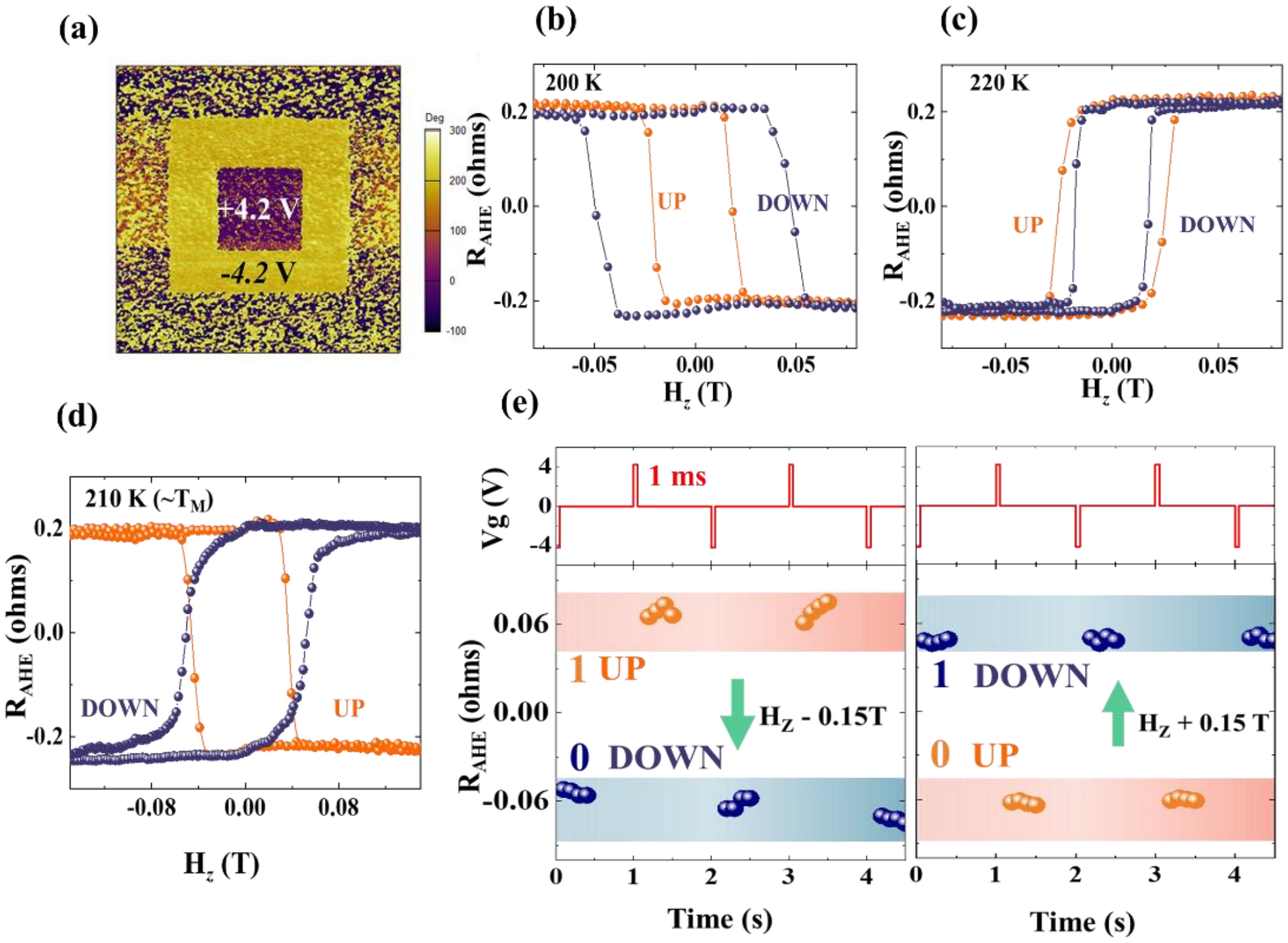

Figure 2 (a) The out-of-plane piezoelectric response of the CCMO/PZT bilayer. The $R_{AHE}$ as a function of a perpendicular field in opposite polarizations at (b) 200 K, (c) 220 K, (d) 210 K. The $V_g$ is +/-4.2 V (100 μs) for up (down) polarized. (e) Pulsed electric fields with width of 100 μs versus the AHE resistance near the $T_M$ point with perpendicular magnetic field of ± 0.15 T, respectively. “1”and “0” denote the positive and negative $R_{AHE}$ respectively.

### 2.3 The mechanism analysis of net magnetization switching

To understand the mechanism of the polarization-induced switching of the magnetization, we performed high-angle annular dark-field (HAADF) and angular bright-field (ABF) measurements to verify what happens to the interface of our heterostructure. As shown in PZT unit cells with c-axis orientation, the polarization vector direction in the PZT is opposite to the direction of Ti atom displacement (with Pb atoms) in the PZT (Figure 3(a)), which is induced by the work function differences between PZT and its bottom electrode [40,41]. Compared with the images of HAADF, ABF images (Figure 3(b)) directly demonstrate the relative displacements of Ti atoms and their oxygen octahedrons. The down polarization induces the downward displacement of Ti atoms displacements' (with oxygen atoms) orientation. Therefore, as illustrated in Figure S5(a), the negative bound charges are accumulated at the interface, which will push oxygen ions from the interface to the CoGd layer and leave a large number of oxygen vacancies at the interfacial PZT. By contrast, the up polarization generates the upward displacement of Ti atoms and relative down movement of oxygen octahedrons, resulting in the positive bound charges accumulated at the interface. As shown in Figure S5(b), this can attract the oxygen ions back into the interface, and decrease the oxygen vacancies at the interface.

To directly demonstrate the oxygen ion migration above, High Resolution Transmission Electron Microscope (HRTEM) images (Figure 3(c)) and the associated EELS spectra were presented at the interface for both polarizations (the red box in Figure 3(c)). The first small peak (529 eV) at the O K edge disappears when the polarization changes from upward to downward (Figure 3(d)), indicating that oxygen vacancies increase at the interface for down polarization [42]. The second peaks (540 eV) at the O K edge are obvious with both polarizations, mainly formed by PZT and partial oxidation in the interfacial CoGd layer. However, when the detect region is about 1.5 nm above the interface (the blue box in Figure 3(c)), EELSs of O-K edges can only be observed with downward polarized (Figure 3(d)), suggesting the oxygen ions increase when the polarization inverts from upward to downward. In addition, in the middle of the CoGd layer, EELSs of O-K edges disappear for both polarizations, and only EELSs of Co-$L_{3,2}$ can be observed (Figure S6), indicating no oxidation in the central region. The results confirm down polarization by electric field can push the oxygen ions from the interface into the nearby CoGd layer, and up polarization can attract them back into the interface. The oxygen ion migrations cause the oxidation reduced/enhanced for upward/downward polarization in the nearby CoGd layer.

.

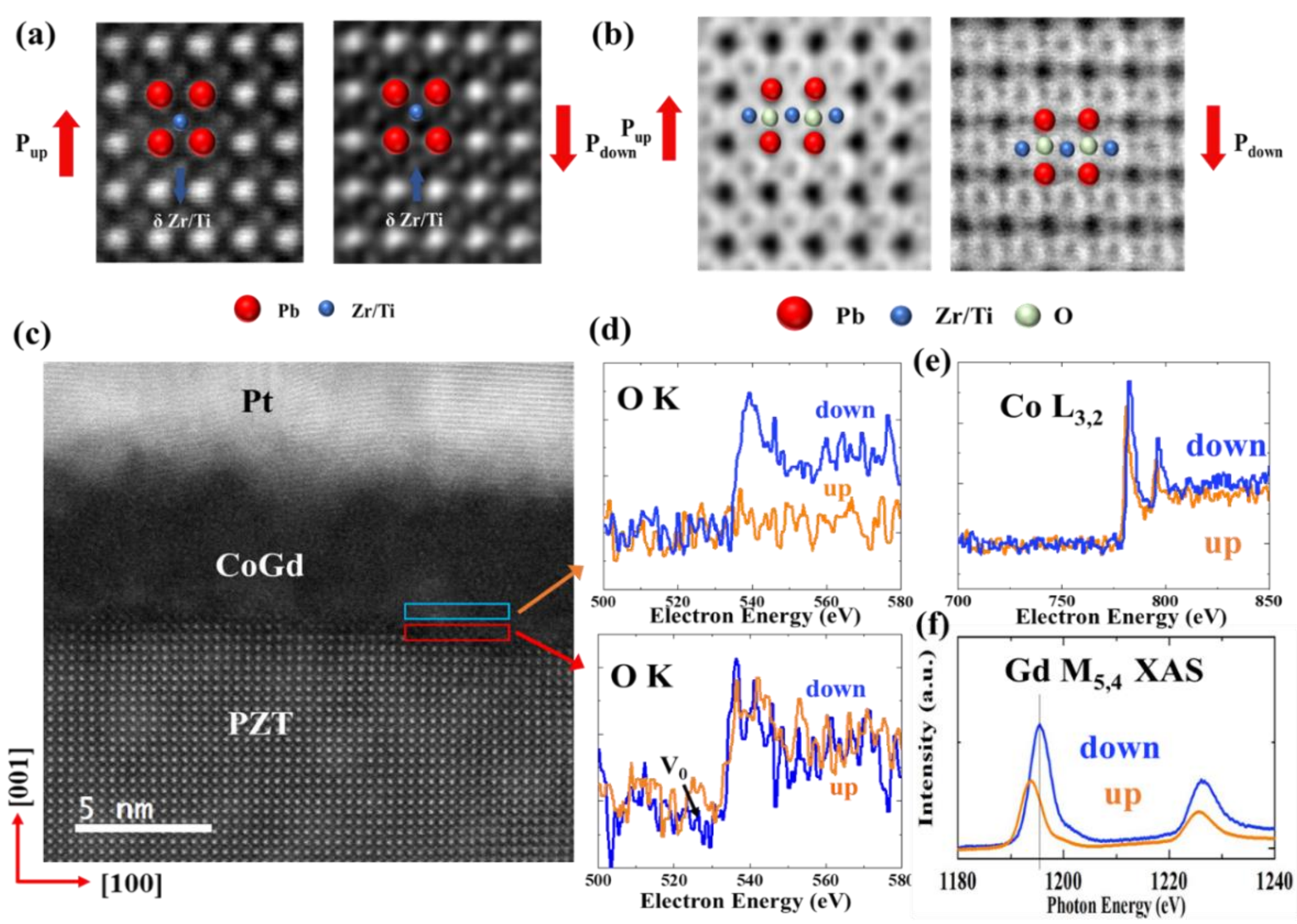

Figure 3 (a) high-angle annular dark-field (HAADF) and (b) angular bright-field (ABF) images of PZT with opposite polarization fields. (c) an HRTEM image of PZT (40 nm)/CoGd (5 nm)/Pt (5 nm) along [100] axis. The red and blue regions represent the interfacial and adjacent areas. (d) EELS spectra of O K and in red and blue regions. The detection width along the [100] axis is ~0.42 nm. (e) EELS spectra of Co $L_{3,2}$ edges with opposite polarization fields. (f) XAS of Gd $M_{5,4}$ edges of PZT/CoGd (0.5 nm)/Pt with opposite polarized directions.

Furthermore, we investigated the corresponding change of oxidation degree for two polarizations by oxygen ion migration in the CoGd layer. The $L_3/L_2$ ratio in EELS spectra of Co-$L_{3,2}$ with down polarization (~1.56) is lower than that with up polarization (~1.98) near the interface (Figure 3(e)), indicating the higher valence states of Co sublattices with downward polarized[42,43]. For Gd sublattices, the peaks of Gd $M_{5,4}$ in XAS have obvious right shift (1194.3 eV to 1196.5 eV, and 1125.3 eV to 1128.6 eV) with polarization changed from upward to downward (Figure 3(f)), suggesting more serious oxidation. Combing with the results of EELSs and XAS, we illustrate the oxygen ion migration for downward polarization can oxide both Co and Gd sublattices

**2.4 Ab initio calculations with a $GdCo_5$ model system**

Finally, to understand the effect of migrated oxygen ion and corresponding change of oxidation degree on the Co and Gd sublattices, we performed ab initio calculations with a representative $GdCo_5$ model system. The surface of the thin film was cleaved across the (100) crystal orientation of the bulk crystalline $GdCo_5$ and terminated with the GdCo layers for its lowest surface energy (Supporting Information S7). Then a supercell $GdCo_5$ layer with a thickness of ~3 nm (Figure S8(a) and (b)) was built to simulate our electric-field control in the PZT/$Co_{77}Gd_{23}$/Pt heterostructure. The Gd/Co magnetization is +7.5 $\mu_B$/ -1.9 $\mu_B$ (Figure S8(c) and (d)), which is in reasonable agreement with the value obtained from XMCD (Supporting Information S9).

To simulate the different oxidation states of the $GdCo_5$ thin film for the upward and downward polarization, interstitial oxygen ions are introduced in the interfacial and interior atomic layers as the experimental observations, respectively. After full structural relaxation, the charge density differences of the $GdCo_5$ alloys are calculated to illustrate the electrons around both Gd and Co atoms are transferred to O atoms for both polarizations (Supporting Information S10).

We further investigated the change of net magnetization for opposite polarizations. For upward polarization, the interfacial oxidation shifts both up and down spin channels of Gd atoms upward. Comparing with that of the spin-down channel, more occupied state electrons were lost for the spin-up channel (Figure 4(a)), resulting in the decrease of the Gd magnetic moment. By contrast, the oxidation of Co atoms and its charge transform mainly influence the spin-up electrons. The shift-up of the spin-up channels gives rise to an enhancement of Co magnetic moment (Figure 4(b)). In this case, the net magnetic moment is calculated to + 0.16 $\mu_B$. This positive magnetic moment indicates the dominance sublattice is still Gd. However, for downward polarization, with serious oxidation, the additional oxygen ions cause interior Gd and Co atoms to decrease occupied state electrons in the same way (Figure 4(c) and (d)). These further decrease the $|M_{Gd}|$ in Gd sublattices while increase the $|M_{Co}|$ in Co-sublattices, resulting in a net magnetization of -0.25 $\mu_B$, which demonstrates the Gd dominance converts to the Co dominance. Therefore, from the ab initio calculations, we conclude that due to the oxygen ion migration, which are from opposite polarization fields, the dominant sublattice switching and the reversal of magnetization can be achieved repeatedly.

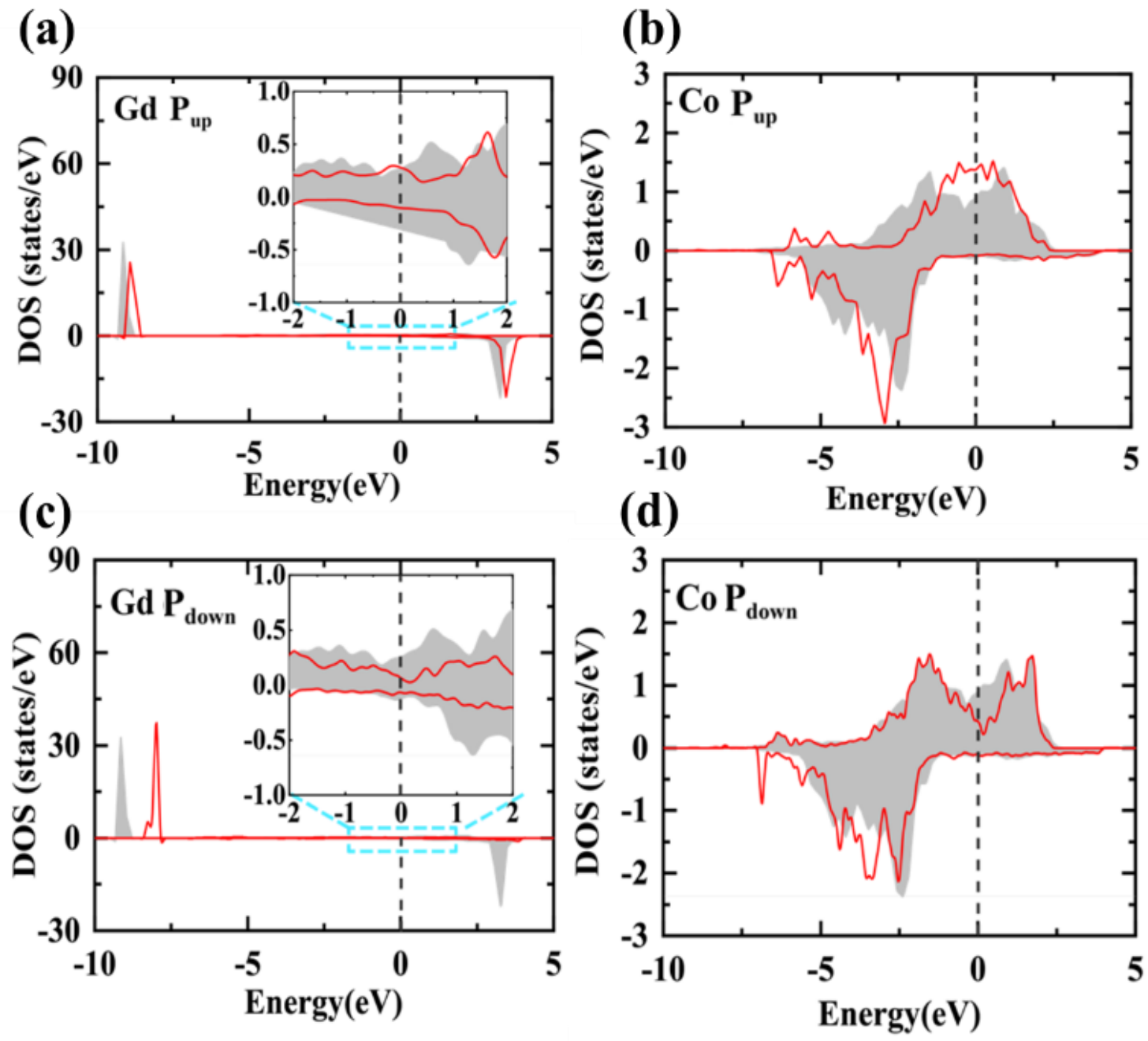

Figure 4 The DFT calculation of Gd (a/c) and Co (b/d) atoms for the upward/downward polarization. (a/b) is from interfacial atoms, while (c/d) is from interior atoms. The gray filled areas present the density of states (DOS) of the Gd or Co atoms in the $GdCo_5$ thin film without polarization. The Fermi level sets at 0 eV.

## 3. Conclusion and Outlook

In conclusion, we demonstrate the switching of perpendicular magnetization in ferrimagnets near the $T_M$ temperature can be controlled by changing the ferroelectric polarization in PZT/CoGd heterostructures. Owing to directly observing the oxygen ion migration and corresponding change of oxidation degree with reversing the polarizations by EELS and XAS, we verified the magnetization control is originated from the oxygen ion migration at the ferroelectric/ferromagnetic interface. Theoretical calculations further confirm it is this migration that change the dominant sublattices in ferrimagnets, inducing the reversible change of net magnetization. Our results provide a method for ferroelectric control of magnetic states and may pave the way for developing solid-state magnetic random memory and logical spintronic devices.

## 4. Experimental Section

Sample preparation: CCMO (20 nm)/PZT (60 nm) were prepared on (100) oriented STO substrate using the pulsed laser deposition technique. The growth conditions of PZT (CCMO) were maintained at 700 (650) °C of substrate temperature, 10 Pa of oxygen pressure, 1.5 (1.2) J/cm$^2$ of laser energy density, and 5 Hz pulse repetition rate. After the deposition, the bilayers were in situ annealed and cooled down to room temperature under an oxygen pressure of 10000 Pa. Then the multilayers of CoGd (3, 4, or 5nm)/Pt (5 nm) were grown using magnetron sputtering. The pressure of the Ar gas was set at 0.8 mTorr for the CoGd/Pt bilayer. The hall-bar devices were patterned using ultraviolet lithography.

Hall effect measurements: AHE measurements at low temperatures were performed using Lakeshore Probe Station, where the vacuum value is below 1.0 $\times 10^{-8}$ mTorr. A Keithley 2403 was used to provide current sources, and a Keithley 2182 was used to detect the hall voltage. In Hall measurement, the applied current is 0.1 mA otherwise noted.

Polarized Methods: Keithley B1500 provided a pulsed voltage to polarize the PZT layer. A square pad in the Hall bar and the CCMO layer were used as top and bottom electrodes. Two probes were used to connect the electrodes with Keithley B1500. A positive (negative) gate voltage of 4.2 V with width of 100 μs was applied to polarize the ferroelectric PZT upward (downward) of the device.

X-ray magnetic circular dichroism (XMCD): In order to ensure the strength of the signals of XMCD of Co and Gd, we reduced the thickness of the protective (Pt) layer down to 1 nm. From the same wafer, samples with opposite polarizations were prepared and further measured in Shanghai Synchrotron Radiation Facility. The measurements were carried out in transmission geometry with a 1 T magnetic field parallel to the incoming X-ray. Two spectra per helicity were recorded in each measurement at different temperatures. The XMCD spectrum was obtained from their differences, while the X-ray absorption spectrum (XAS) spectrum was calculated from the sum of two helicity spectra. In addition, the magnetizations were calculated by $M_Z = -\mu_B/\hbar(L_Z +$ $2S_Z$), where $\mu_B$ is the Bohr magneton, $\hbar$ is the reduced Planck constant, $L_Z$ is the z-component of the orbital angular momentum, and $S_Z$ is the z-component of the spin angular momentum. For Co, $L_Z = 4\hbar n_h q/3r$, $q$ is the total XMCD integral of Co $L_2$ and $L_3$ peaks, and $r$ is the total integral over the XAS after background correction, $n_h = 2.5$. $S_Z = \hbar n_h(3p - 2q)/r$, where $p$ is the XMCD integral of the $L_3$ peak. For Gd, $L_Z = 2\hbar n_h q/r$, where $n_h = 7$ and $q$ is the total XMCD integral of Gd $M_4$ and $M_5$ peaks, and r is the total integral over the XAS after background correction. $S_Z = \frac{S_{eff}}{2+\frac{6T_Z^{free}}{S_Z^{free}}}$, where $S_{eff} = \hbar n_h(5p - 3q)$/r and $\frac{S_Z^{free}}{T_Z^{free}} = -3.466/0.01$.

**Acknowledgments**

This work was supported by the National Key Research and Development Program of China (Grant No. 2022YFA1405100), by Beijing Natural Science Foundation Grant No. 2212048. This work was also supported by the NSFC Grant No. 12241405, 12304154, 52025025, 52072400, 52250402 and Natural Science Foundation of Zhejiang Provincial Grant No. LY22E020012. And the project was sponsored by the Chinese Academy of Sciences, Grant No. XDB28000000, and XDPB44000000 as well. We also thank the support of the Shanghai Synchrotron Radiation Facility.

**Conflict of interest**

The authors declare no conflicts of interest.

**Data Availability Statement**

The data that support the findings of this study are available from the corresponding author upon reasonable request.

**References**

1. P. Borisov, A. Hochstrat, X. Chen, W. Kleemann, C. Binek, Magnetoelectric Switching of Exchange Bias. *Phys. Rev. Lett.* **2005,** 94, 117203.
2. S. S. P. Parkin, M. Hayashi, L. Thomas, Magnetic Domain-Wall Racetrack Memory. *Science* **2008,** 320, 190–194
3. E. Y. Tsymbal, Electric toggling of magnets. *Nat. Mater.* **2012,** 11, 12–13.
4. W. Wang, M. Li, S. Hageman, C. L. Chien, Electric-field-assisted switching in magnetic tunnel&junctions. *Nat. Mater.* **2011,** 11, 64–68.
5. F. Matsukura, Y. Tokura, H. Ohno, Control of magnetism by electric fields. Nat. *Nanotechnol.* **2015,** 10, 209–220.
6. S. Zhang, Y. Zhao, X. Xiao, Y. Wu, S. Rizwan, L. Yang, P. Li, J. Wang, M. Zhu, H. Zhang, X. Jin, X. Han, Giant electrical modulation of magnetization in $Co_{40}Fe_{40}B_{20}/Pb(Mg_{1/3}Nb_{2/3})_{0.7}Ti_{0.3}O_3$(011) heterostructure. *Sci. Rep.* **2015,** 4, 3727.
7. B. Zhang, H.-L. Wang, J. Cao, Y.-C. Li, M.-Y. Yang, K. Xia, J.-H. Zhao, K.-Y. Wang, Control of magnetic anisotropy in epitaxial $Co_2MnAl$ thin films through piezo-voltage-induced strain. *J. Appl. Phys.* **2019,** 125, 82503.
8. B. Zhang, K.-K. Meng, M.-Y. Yang, K. W. Edmonds, H. Zhang, K.-M. Cai, Y. Sheng, N. Zhang, Y. Ji, J.-H. Zhao, H.-Z. Zheng, K.-Y. Wang, Piezo Voltage Controlled Planar Hall Effect Devices. *Sci. Rep.* **2016,** 6, 28458.
9. Y. Li, W. Luo, L. Zhu, J. Zhao, K. Wang, Voltage manipulation of the magnetization reversal in Fe/n-GaAs/piezoelectric heterostructure. *J. Magn. Magn. Mater.* **2015,** 375, 148–152.
10. Q. He, Y. H. Chu, J. T. Heron, S. Y. Yang, W. I. Liang, C. Y. Kuo, H. J. Lin, P.

Yu, C. W. Liang, R. J. Zeches, W. C. Kuo, J. Y. Juang, C. T. Chen, E. Arenholz, A. Scholl, R. Ramesh, Electrically controllable spontaneous magnetism in nanoscale mixed phase multiferroics. *Nat. Commun.* **2011,** 2, 225.

11. F. Zhang, H. Fan, B. Han, Y. Zhu, X. Deng, D. Edwards, A. Kumar, D. Chen, X. Gao, Z. Fan, B. J. Rodriguez, Boosting Polarization Switching-Induced Current Injection by Mechanical Force in Ferroelectric Thin Films. *ACS Appl. Mater. Interfaces* **2021,** 13, 26180–26186.

12. D. Yi, P. Yu, Y.-C. Chen, H.-H. Lee, Q. He, Y.-H. Chu, R. Ramesh, Tailoring magnetoelectric coupling in $BiFeO_3/La_{0.7}Sr_{0.3}MnO_3$ heterostructure through the interface engineering. *Adv. Mater.* **2019,** 31, 180633.

13. N. A. Spaldin, R. Ramesh, Advances in magnetoelectric multiferroics. *Nat. Mater.* **2019,** 18, 203–212.

14. J. T. Heron, J. L. Bosse, Q. He, Y. Gao, M. Trassin, L. Ye, J. D. Clarkson, C. Wang, J. Liu, S. Salahuddin, D. C. Ralph, D. G. Schlom, J. Íñiguez, B. D. Huey, R. Ramesh, Deterministic switching of ferromagnetism at room temperature using an electric field. *Nature* **2014,** 516, 370–373.

15. X. Jiang, L. Gao, J. Z. Sun, S. S. P. Parkin, Temperature dependence of current-induced magnetization switching in spin valves with a ferrimagnetic CoGd free layer. *Phys. Rev. Lett.* **2006,** 97, 1–4.

16. C. D. Stanciu, A. V Kimel, F. Hansteen, A. Tsukamoto, A. Itoh, A. Kirilyuk, Th. Rasing, Ultrafast spin dynamics across compensation points in ferrimagnetic GdFeCo: The role of angular momentum compensation. *Phys. Rev. B* **2006,** 73, 22040.

17. L. Caretta, M. Mann, F. Büttner, K. Ueda, B. Pfau, C. M. Günther, P. Hessing, A. Churikova, C. Klose, M. Schneider, D. Engel, C. Marcus, D. Bono, K. Bagschik, S. Eisebitt, G. S. D. Beach, Fast current-driven domain walls and small skyrmions in a compensated ferrimagnet. *Nat. Nanotechnol.* **2018,** 13, 1154–1160.

18. K. Cai, Z. Zhu, J. M. Lee, R. Mishra, L. Ren, S. D. Pollard, P. He, G. Liang, K.

L. Teo, H. Yang, Ultrafast and energy-efficient spin–orbit torque switching in compensated ferrimagnets. *Nat. Electron.* **2020,** 3, 37–42.

19. A. Fert, R. Ramesh, V. Garcia, F. Casanova, M. Bibes, Electrical control of magnetism by electric field and current-induced torques. *Rev. Mod. Phys.* **2024,** 96, 015005.

20. S. K. Kim, G. S. D. Beach, K. J. Lee, T. Ono, T. Rasing, H. Yang, Ferrimagnetic spintronics. *Nat. Mater.* **2022,** 21, 24–34.

21. K.-J. Kim, S. K. Kim, Y. Hirata, S.-H. Oh, T. Tono, D.-H. Kim, T. Okuno, W. S. Ham, S. Kim, G. Go, Y. Tserkovnyak, A. Tsukamoto, T. Moriyama, K.-J. Lee, T. Ono, Fast domain wall motion in the vicinity of the angular momentum compensation temperature of ferrimagnets. *Nat. Mater.* **2017,** 16, 1187–1192.

22. K. Ueda, M. Mann, P. W. P. de Brouwer, D. Bono, G. S. D. Beach, Temperature dependence of spin-orbit torques across the magnetic compensation point in a ferrimagnetic TbCo alloy film. *Phys. Rev. B* **2017,** 96, 064410.

23. R. Mishra, J. Yu, X. Qiu, M. Motapothula, T. Venkatesan, H. Yang, Anomalous Current-Induced Spin Torques in Ferrimagnets near Compensation. *Phys. Rev. Lett.* **2017,** 118, 167201.

24. M. Huang, M. U. Hasan, K. Klyukin, D. Zhang, D. Lyu, P. Gargiani, M. Valvidares, S. Sheffels, A. Churikova, F. Büttner, J. Zehner, L. Caretta, K. Y. Lee, J. Chang, J. P. Wang, K. Leistner, B. Yildiz, G. S. D. Beach, Voltage control of ferrimagnetic order and voltage-assisted writing of ferrimagnetic spin textures. *Nat. Nanotechnol.* **2021,** 16, 981–988.

25. J. Wang, M. Li, C. Li, R. Tang, M. Si, G. Chai, J. Yao, C. Jia, C. Jiang, Piezostrain-controlled magnetization compensation temperature in ferrimagnetic GdFeCo alloy films. *Phys. Rev. B* **2023,** 107, 184424.

26. X. Feng, Z. Zheng, Y. Zhang, Z. Zhang, Y. Shao, Y. He, X. Sun, L. Chen, K. Zhang, P. Khalili Amiri, W. Zhao, Magneto-ionic Control of Ferrimagnetic Order by Oxygen Gating. *Nano Lett.* **2023,** 23, 4778–4784.

27. P. Hansen, S. Klahn, C. Clausen, G. Much, K. Witter, Magnetic and magneto-

optical properties of rare-earth transition-metal alloys containing Dy, Ho, Fe, Co. *J. Appl. Phys.* **1991,** 69, 3194–3207.

28. L. W. Martin, A. M. Rappe, Thin-film ferroelectric materials and their applications. Nat. Rev. Mater. **2016,** 2, 16087.
29. W. Eerenstein, N. D. Mathur, J. F. Scott, Multiferroic and magnetoelectric materials. Nature **2006,** 442, 759–765.
30. W. Ci, P. Wang, W. Xue, H. Yuan, X. Xu, Engineering Ferroelectric-/Ion-Modulated Conductance in 2D vdW $CuInP_2S_6$ for Non-Volatile Digital Memory and Artificial Synapse. Adv. Funct. Mater. **2024,** 2316360.
31. Z. Q. Liu, H. Chen, J. M. Wang, J. H. Liu, K. Wang, Z. X. Feng, H. Yan, X. R. Wang, C. B. Jiang, J. M. D. Coey, A. H. MacDonald, Electrical switching of the topological anomalous Hall effect in a non-collinear antiferromagnet above room temperature. Nat. Electron. **2018,** 1, 172–177.
32. H. Yan, Z. Feng, S. Shang, X. Wang, Z. Hu, J. Wang, Z. Zhu, H. Wang, Z. Chen, H. Hua, W. Lu, J. Wang, P. Qin, H. Guo, X. Zhou, Z. Leng, Z. Liu, C. Jiang, M. Coey, Z. Liu, A piezoelectric, strain-controlled antiferromagnetic memory insensitive to magnetic fields. Nat. Nanotechnol. **2019,** 14, 131–136.
33. N. Lei, T. Devolder, G. Agnus, P. Aubert, L. Daniel, J.-V. Kim, W. Zhao, T. Trypiniotis, R. P. Cowburn, C. Chappert, D. Ravelosona, P. Lecoeur, Strain-controlled magnetic domain wall propagation in hybrid piezoelectric/ferromagnetic structures. Nat. Commun. **2013,** 4, 1378.
34. B. Cui, C. Song, H. Mao, Y. Yan, F. Li, S. Gao, J. Peng, F. Zeng, F. Pan, Manipulation of Electric Field Effect by Orbital Switch. Adv. Funct. Mater. **2016,** 26, 753–759.
35. B. F. Vermeulen, F. Ciubotaru, M. I. Popovici, J. Swerts, S. Couet, I. P. Radu, A. Stancu, K. Temst, G. Groeseneken, C. Adelmann, K. M. Martens, Ferroelectric Control of Magnetism in Ultrathin $HfO_2$\Co\Pt Layers. ACS Appl. Mater. Interfaces **2019,** 11, 34385–34393.
36. I. Stolichnov, S. W. E. Riester, H. J. Trodahl, N. Setter, A. W. Rushforth, K. W.

Edmonds, R. P. Campion, C. T. Foxon, B. L. Gallagher, T. Jungwirth, Non-volatile ferroelectric control of ferromagnetism in (Ga,Mn)As. Nat. Mater. **2008,** 7, 464–467.

37. X. Huang, X. Chen, Y. Li, J. Mangeri, H. Zhang, M. Ramesh, H. Taghinejad, P. Meisenheimer, L. Caretta, S. Susarla, R. Jain, C. Klewe, T. Wang, R. Chen, C.-H. Hsu, I. Harris, S. Husain, H. Pan, J. Yin, P. Shafer, Z. Qiu, D. R. Rodrigues, O. Heinonen, D. Vasudevan, J. Íñiguez, D. G. Schlom, S. Salahuddin, L. W. Martin, J. G. Analytis, D. C. Ralph, R. Cheng, Z. Yao, R. Ramesh, Manipulating chiral spin transport with ferroelectric polarization. Nat. Mater. 2024, doi: 10.1038/s41563-024-01854-8.

38. K. Cai, M. Yang, H. Ju, S. Wang, Y. Ji, B. Li, K. W. Edmonds, Y. Sheng, B. Zhang, N. Zhang, S. Liu, H. Zheng, K. Wang, Electric field control of deterministic current-induced magnetization switching in a hybrid ferromagnetic/ferroelectric structure. Nat. Mater. **2017,** 16, 712–716.

39. V. López-Flores, N. Bergeard, V. Halté, C. Stamm, N. Pontius, M. Hehn, E. Otero, E. Beaurepaire, C. Boeglin, Role of critical spin fluctuations in ultrafast demagnetization of transition-metal rare-earth alloys. *Phys. Rev. B* 2013, 87, 214412.

40. P. Gao, C. T. Nelson, J. R. Jokisaari, S.-H. Baek, C. W. Bark, Y. Zhang, E. Wang, D. G. Schlom, C.-B. Eom, X. Pan, Revealing the role of defects in ferroelectric switching with atomic resolution. *Nat. Commun.* **2011,** 2, 591.

41. M. Z. Wang, H. J. Feng, C. X. Qian, J. He, J. Feng, Y. H. Cao, K. Yang, Z. Y. Deng, Z. Yang, X. Yao, J. Zhou, S. (Frank) Liu, X. C. Zeng, $PbTiO_3$ as Electron-Selective Layer for High-Efficiency Perovskite Solar Cells: Enhanced Electron Extraction via Tunable Ferroelectric Polarization. *Adv. Funct. Mater.* **2019,** 29, 1806427.

42. Q. Zhang, X. He, J. Shi, N. Lu, H. Li, Q. Yu, Z. Zhang, L. Q. Chen, B. Morris, Q. Xu, P. Yu, L. Gu, K. Jin, C. W. Nan, Atomic-resolution imaging of electrically induced oxygen vacancy migration and phase transformation in $SrCoO_{2.5-\sigma}$. *Nat.*

*Commun.* **2017,** 8, 1–6.

43. Z. L. Wang, J. S. Yin, Y. D. Jiang, EELS analysis of cation valence states and oxygen vacancies in magnetic oxides. in *Micron,* **2000,** vol. 31 571–580.